\documentclass[notitlepage, final, 10pt, pra, aps, floatfix, twocolumn, letter, english, superscriptaddress]{revtex4-2}
\pdfoutput=1

\usepackage[T1]{fontenc}
\usepackage[utf8]{inputenc}
\usepackage[english]{babel}
\usepackage{lmodern}
\usepackage{amsfonts}
\usepackage{amssymb}
\usepackage{amsmath}
\usepackage{bm}
\usepackage{bbm}
\usepackage{cases}
\usepackage{graphicx}
\usepackage{subfigure}
\usepackage{xcolor}
\definecolor{navyblue}{rgb}{0.0, 0.0, 0.5}
\usepackage{latexsym}
\usepackage{booktabs}
\usepackage[protrusion=true, expansion=true]{microtype}
\usepackage[colorlinks=true, breaklinks=false, linkcolor=navyblue, urlcolor=navyblue, citecolor=navyblue]{hyperref}
\usepackage{float}
\usepackage{mathtools}
\usepackage{braket}
\usepackage{siunitx}
\newcommand{\ham}{\hat H}
\newcommand{\aop}{\hat a}
\newcommand{\aopd}{\hat a^\dagger}
\newcommand{\bop}{\hat b}
\newcommand{\bopd}{\hat b^\dagger}
\newcommand{\noa}{\aopd \aop}
\newcommand{\nob}{\bopd \bop}
\newcommand{\tls}{\text{TLS}}
\newcommand{\dens}{\hat \rho}
\makeatother

\begin{document}

\title{Two-tone spectroscopy for the detection of two-level systems in superconducting qubits}
\author{Olli Mansikkamäki}
\affiliation{Nordita,
Stockholm University and KTH Royal Institute of Technology,
Hannes Alfvéns väg 12, SE-106 91 Stockholm, Sweden}

\author{Alexander Tyner}
\affiliation{Nordita,
Stockholm University and KTH Royal Institute of Technology,
Hannes Alfvéns väg 12, SE-106 91 Stockholm, Sweden}
\affiliation{Department of Physics, University of Connecticut, Storrs, Connecticut 06269, USA}

\author{Alexander Bilmes}
\affiliation{Google Research, Mountain View, CA, USA.}

\author{Ilya Drozdov}
\affiliation{Google Research, Mountain View, CA, USA.}
\affiliation{Department of Physics, University of Connecticut, Storrs, Connecticut 06269, USA}

\author{Alexander Balatsky}

\affiliation{Nordita, 
Stockholm University and KTH Royal Institute of Technology,
Hannes Alfvéns väg 12, SE-106 91 Stockholm, Sweden}

\affiliation{Department of Physics, University of Connecticut, Storrs, Connecticut 06269, USA}

\begin{abstract}
    Two-level systems (TLS) of unclear physical origin are a major contributor to decoherence in superconducting qubits. The interactions of individual TLS with a qubit can be detected via various spectroscopic methods, most of which have relied on the tunability of the qubit frequency. We propose a novel method that requires only a microwave drive and dispersive readout, and thus also works fixed-frequency qubits. The proposed two-tone spectroscopy involves a microwave pulse of varying frequency and length to excite TLSs of unknown frequencies, followed by a second pulse at the qubit frequency. TLS parameters can be estimated from the qubit population as a function of the first pulse frequency and length.  
\end{abstract}

\maketitle

\section{Introduction}
Various attempts at building a quantum computer using superconducting qubits technology have been made recently. Although significant improvements  have been made in this approach, the further progress on this path is increasing limited by the noise and decoherence of these devices.  A major role to the decoherence can be attributed to various two-level systems (TLS) coupled to the qubit building blocks of the device~\cite{kjaergaard_superconducting_2020, wang_surface_2015, gambetta_investigating_2017, barends_coherent_2013, muller_towards_2019, klimov_fluctuations_2018, burnett_decoherence_2019, schlor_correlating_2019, spiecker_two-level_2023, thorbeck_readout-induced_2024}. While the exact physical origin of the TLS is still unclear, they are often though to exist in the amorphous interfaces contained in the circuitry of the physical qubit. Thus, the phenomenological model~\cite{anderson_anomalous_1972, phillips_tunneling_1972} developed to explain the low-temperature specific heat and thermal conductivity of amorphous solids is still used in noise in the context of superconducting qubits.

While the original model describes an ensemble of TLSs, we are of opinion that the proper characterization, control and ultimately, rectification of the effects of TLSs would require an identification and understanding of a single TLS as a physical entity, its coupling to qubit. Therefore the focus of this paper is on spectroscopy that identifies the TLS and allows one to estimate their parameters {\it one at a time}.

Individual TLSs have been previously experimentally detected and even coherently controlled~\cite{lisenfeld_measuring_2010, abdurakhimov_driven-state_2020, abdurakhimov_identification_2022, chen_phonon_2023}. Those experiments usually involve tuning the qubit frequency to match that of the TLS. However, one could conceivably opt to use fixed-frequency qubits in their device, as they are less sensitive to flux noise and often have longer lifetimes than their tunable counterparts~\cite{paraoanu_microwave-induced_2006, rigetti_fully_2010, de_groot_selective_2010, krantz_quantum_2019}. One can presume the fixed-frequency qubits to be affected by at least some of the types of TLS that afflict tunable ones. Therefore, we propose a method for the detection of strongly coupled {\it individual} TLS that requires no frequency tuning, and relies only on XY-control and dispersive readout of the qubit population. 

\begin{figure}
    \centering
    \includegraphics[width=\columnwidth]{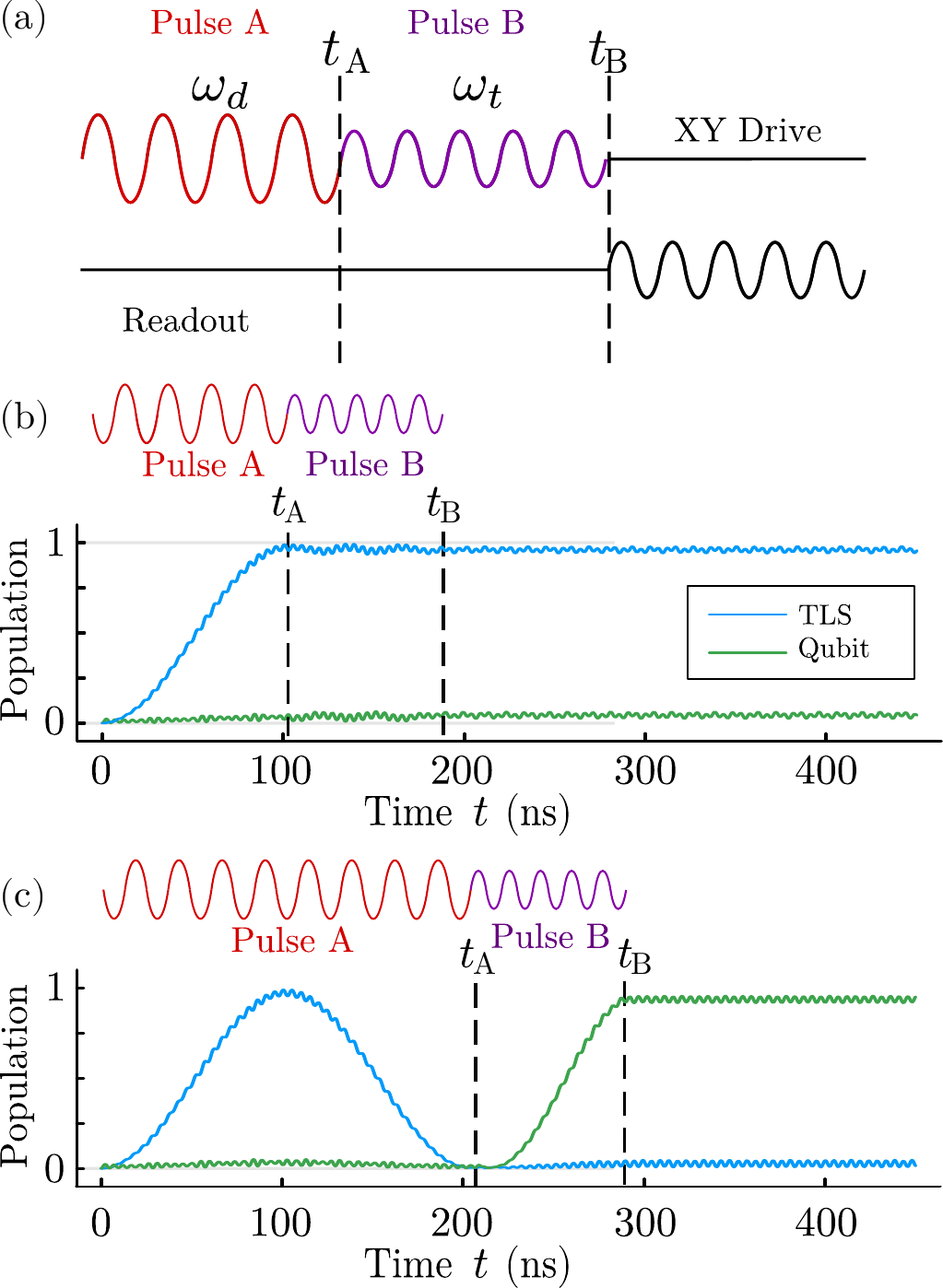}
    \caption{(a) The pulse sequence for exciting and probing strongly coupled TLS in the vicinity of the transmon frequency. (b), (c) The population of a TLS (blue) and the transmon (green) at when driven at the TLS frequency for different pulse A lengths $t_A$. (b) The TLS population is near maximum at $t_A$, and pulse B has little effect on the transmon population. (c) The TLS population is near zero at $t_A$, and pulse B excites the transmon population to one.}
    \label{fig:fig1}
\end{figure}

Similarly to microwave-only two-qubit gates~\cite{rigetti_fully_2010, chow_simple_2011, poletto_entanglement_2012}, we can utilize the fact that a TLS can be excited by driving the qubit~\cite{lisenfeld_measuring_2010, chen_phonon_2023}. Assuming we have no prior knowledge of the frequency of the TLS, we must drive the qubit with a range of frequencies. Since we also do not know how long the pulse needs to be to get the TLS to its excited state, the length of the pulse must also be varied. Knowing the population of the TLS as a function of pulse lengths and frequencies, one could extract parameters of the TLS, such as its frequency and the strength of its coupling to the qubit. Unfortunately, it is not possible to directly measure the population of the TLS. Due to the coupling, however, the frequency of the qubit depends on the state of the TLS. A precise $\pi$-pulse calibrated at the qubit frequency when the TLS are not excited will also result in different qubit population depending on the state of the TLS. That is, we first drive the qubit at some frequency $\omega_d$ for time $t_A$; see Fig.~\ref{fig:fig1} (a). This is followed by a $\pi$-pulse at the qubit frequency and the measurement of the qubit population. The two-pulse sequence is then repeated for a range of frequencies and pulse lengths. The resulting ($\omega, t$)-map contains valuable information on the TLS parameters, and represents an example of higher order spectroscopy akin to two dimensional NMR. 

Each ($\omega, t$)-map is associated with a well defined set of labels in terms of the parameters of the qubit and the TLSs. Extraction of the TLS parameters can therefore be accurately and efficiently accomplished using convolutional neural networks. We benchmark this approach through construction of a neural network for the identification of TLS frequencies in the presence of multiple TLSs.

While the method should work for a variety of superconducting qubit designs, we will focus on the transmon~\cite{koch_charge-insensitive_2007, paik_observation_2011, barends_coherent_2013} due to its popularity as a platform. The transmon is an anharmonic oscillator, whose two lowest states are usually operated as a qubit. However, due to its relatively weak anharmonicity, the second excited level needs to be taken into account in order to avoid leakage outside the computational subspace. This being a purely theoretical study, our choice of focusing on the transmon is largely seen in the typical parameter ranges and the sign of the anharmonicity. 

The article is organised as follows. In Sec.~\ref{sec:model}, we introduce the Hamiltonian we use to model the coherent behavior of the transmon-TLS system, and the Lindblad master equation we use for decoherence. We begin Sec.~\ref{sec:maps} by describing the TLS detection and probing method. This is followed by numerically simulated examples of the behavior of the system with and without decoherence in Sec.~\ref{sec:coherent} and Sec.~\ref{sec:decoherent}, respectively. The case of multiple TLSs and how to deal with them using machine learning is discussed in Sec.~\ref{sec:ML}. Conclusions and outlook are presented in Sec.~\ref{sec:conclusions}.

\section{Transmon and two-level systems}\label{sec:model}
The first few energy levels of a transmon can be reasonably accurately be described with the Hamiltonian~\cite{krantz_quantum_2019}
\begin{equation}\label{eq:transmon}
\ham_q / \hbar = \omega_q \noa - \frac{U}{2} \aopd \aopd \aop \aop, 
\end{equation}
written in the basis of bosonic annihilation and creation operators $\aop$ and $\aopd$. Here $\omega_q$ is the bare frequency of the transmon, and $U$ its anharmonicity. Any transmon will likely have at least a few TLSs with frequencies $\omega_k$ close enough to to the transmon frequency, and with strong enough couplings $g_k$ for the TLSs to significantly affect the performance of the transmon. We model these with the Hamiltonian
\begin{equation}
\ham_\tls / \hbar = \sum_k \omega_k \bopd_k \bop_k + \sum_k g_k (\aopd \bop_k + \aop \bopd_k),
\end{equation}
where $\bop$ and $\bopd$ are hard-core boson ($\bopd \bopd = 0$) annihilation and creation operators. Note that the Hamiltonian does not include direct interaction between the TLSs, which choice is supported by experiment~\cite{chen_phonon_2023}. 

The microwave drive is described by the Hamiltonian
\begin{equation}\label{eq:hamd}
\ham_d / \hbar = A(t) \cos (t \omega_d(t))\left(\aop + \aopd + \sum_k \lambda_k (\bop + \bopd)\right).
\end{equation}
Driving the transmon can also directly affect the populations of the TLSs. Here we assume the effect to be similar to the effect on the transmon, although at magnitude $\lambda_k$. 

The unitary time-evolution of the transmon-TLS system is modeled with the Hamiltonian
\begin{equation}
    \ham = \ham_q + \ham_\tls + \ham_d.
\end{equation}
We model the effect of the environment not captured by the above Hamiltonian with the Lindblad master equation
\begin{align}\label{eq:lindblad}
    \frac{d \dens}{d t} &= -\frac{i}{\hbar} [\ham, \dens] + \frac{\gamma_q}{2}(2 \aop \dens \aopd - \noa \dens - \dens \noa) \nonumber \\
    &+ \frac{\kappa_q}{2} (2 \noa \dens \noa - (\noa)^2 \dens - \dens (\noa)^2) \\
    &+ \sum_k \frac{\gamma_k}{2}(2 \bop \dens \bopd - \nob \dens - \dens \nob) \nonumber \\
    &+ \sum_k \frac{\kappa_k}{2} (2 \nob \dens \nob - \nob \dens - \dens \nob).
\end{align}
Here the second term describes the dissipation from the transmon to its environment, excluding the TLSs. Similarly, the third term describes pure dephasing of the transmon. The last two terms describe the dissipation and dephasing of the TLSs.

Typically in transmon based devices~\cite{koch_charge-insensitive_2007, ma_dissipatively_2019, arute_quantum_2019, paik_observation_2011, morvan_formation_2022, zhu_observation_2022, chen_phonon_2023}, the frequencies are of the order $\omega_q / 2 \pi \sim \SIrange{3}{8}{\giga\hertz}$ and their anharmonicities $U / 2\pi \sim \SIrange{150}{250}{\mega\hertz}$. The parameters of the TLSs vary between devices and indeed in time. Using the table of the TLS parameters in Ref.~\cite{chen_phonon_2023} as a guide, we assume the coupling strengths of individually detectable TLSs fit roughly in the range $g_k / 2\pi \sim \SIrange{5}{50}{\mega\hertz}$ and their dissipation times $T_{1, k} \sim \SIrange{0.5}{20}{\micro\second}$. Scant expertimental data for the pure dephasing times $T_{\varphi,k}$ exists, and what little there is seems to suggest that they can range from similar to the dissipation times to significantly higher, even to the point of being negligible~\cite{lisenfeld_measuring_2010, lisenfeld_decoherence_2016, chen_phonon_2023}.

\section{($\omega, t$)-maps}\label{sec:maps}
For the sake of simplicity, we will assume there to be a single TLS until noted otherwise. We also assume that the only measurable quantity available to us is the population of the first excited level of the transmon $P_q(t) = \bra{\psi(t)} (\ket{10}\bra{10} + \ket{11}\bra{11}) \ket{\psi(t)}$. Here in the Fock state $\ket{nm}$, $n$ and $m$ are the occupation numbers of the transmon and the TLS, respectively. This limitation in observables does not, however, preclude the probing of the state of a TLS without frequency tuning. We can exploit the fact that the driving frequencies resulting in the transitions $\ket{00} \to \ket{10}$ and $\ket{01} \to \ket{11}$ differ when the second excited level of the transmon is taken into account. In the limit $A, g \ll \omega_q$, this difference is approximately
\begin{equation}\label{eq:shift}
    \delta \omega = \frac{4 g_k^2}{U - \Delta_k} + \frac{A^2 \Delta_k}{g_k^2}.
\end{equation}
As a result, a $\pi$-pulse calibrated for the $\ket{10}$ transition will produce a notably diminished effect for the $\ket{11}$ transition. Noting that the difference is not all that large, pulse B needs to be quite precise. We use a relatively long $\pi$-pulse of $t_\pi = \SI{100}{\nano\second}$ with a simple rectangular envelope for the sake of numerical expediency. It is possible that with e.g. a Gaussian envelope or a more involved pulse scheme~\cite{motzoi_simple_2009, chow_optimized_2010, lucero_reduced_2010}, the length could be reduced. 

{\em Protocol}. We outline the set up to generate ($\omega, t$) maps: 
 \begin{itemize}
    \item First, prepare the system in its ground state $\ket{00}$, e.g., by waiting until the population is fully dissipated. 
    \item Second, drive the transmon with the drive frequency $\omega_d$ until time $t_A$. We label this pulse A.
    \item Third, perform a $\pi$-pulse of length $t_\pi$ at the measured frequency $\tilde \omega_q$. We label this pulse B.
    \item Last, measure the transmon population. Due to the changing TLS population, the post-pulse B population of the transmon at time $t_B = t_A + t_\pi$ will vary depending on the time $t_A$; see Fig.~\ref{fig:fig1} (c). 
 \end{itemize} 
These steps are then repeated for a range of drive frequencies $\omega_d$ and lengths $t_A$ of the pulse A. The parameters of the TLSs, i.e., the frequencies $\tilde \omega_k$, the coupling strengths $g_k$, and the dissipation and pure dephasing times $T_{1, k} = 1 / \gamma_k$ and $T_{\varphi, k} = 2 / \kappa_k$, are all visible in the resulting data, as shown in the next sections.

\subsection{Coherent transmon-TLS coupling}\label{sec:coherent}
\begin{figure}
    \centering
    \includegraphics[width = \columnwidth]{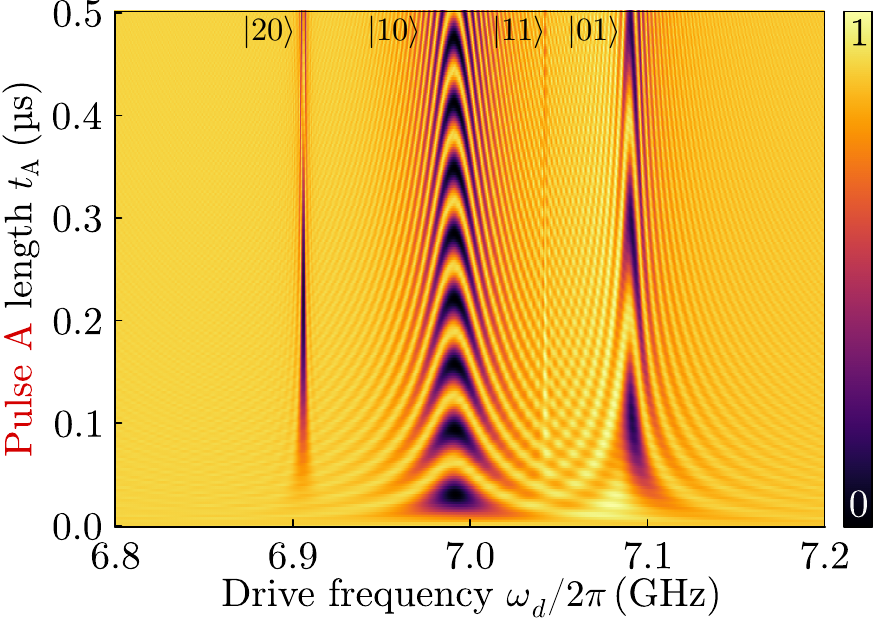}
    \caption{(a) The population $P_q(t_2)$ of the first excited level of the transmon after pulse B as a function of the pulse A frequency $\omega_d$ and length $t_A$. The labels in the middle mark the state $\ket{nm}$ of the $\ket{00} \leftrightarrow \ket{nm}$ transition represented by the feature. }
    \label{fig:fig2}
\end{figure}

Let us start by ignoring the dissipation and dephasing. An example $(\omega, t)$-map produced via a numerical simulation of the procedure described above is shown in Fig.~\ref{fig:fig2}(a). With our modeling of the transmon limited to the three lowest levels, there are two possible transitions from the ground state to the excited levels of the transmon $\ket{10}$ and $\ket{20}$. The latter of these is also rendered visible by pulse B, and is located approximately at
\begin{equation}
    \omega_d = \omega_q - \frac{U}{2} - \frac{g_k^2}{U - \Delta_k}.
\end{equation}

A single TLS will have three possible transitions: from the ground state to $\ket{01}$, $\ket{11}$, and $\ket{21}$. The last of these will usually be slow compared to typical dissipation times, so we shall ignore it here. Out of the other two, let us look at the $\ket{01}$ transition first. This corresponds to the two-qubit cross-resonance gate~\cite{rigetti_fully_2010, chow_simple_2011}. When the transmon is driven at the shifted frequency of the TLS
\begin{equation}\label{eq:drive01}
    \omega_d = \omega_k - \frac{g^2}{\Delta_k},
\end{equation}
its population undergoes Rabi oscillation at the frequency
\begin{equation}\label{eq:rabi01}
    \Omega_\text{01} = \frac{g_k A}{\tilde \Delta_k} + \lambda_k A,
\end{equation}
where $\tilde \Delta_k = \Delta_k + 2 g^2 / \Delta_k$
Note that this feature on it's own is enough for the estimation of the bare frequency $\omega_k$ of the TLS, and the strength $g_k$ of it's coupling of the transmon. 

Another feature of interest is the $\ket{11}$ transition, which corresponds to the bSWAP two-qubit gate~\cite{poletto_entanglement_2012}. In this case, the transmon is driven at a frequency close to the midpoint between the $\ket{10}$ and $\ket{01}$ transition frequencies,
\begin{equation}\label{eq:drive11}
    \omega_d = \frac{\omega_q + \omega_k}{2} + \frac{g^2}{U - \tilde \Delta_k}.
\end{equation}
This results in the population of both oscillating at the same frequency, approximately given by~\cite{poletto_entanglement_2012}
\begin{equation}\label{eq:rabi11}
    \Omega_{11} = \frac{2 A^2 g (U + \lambda^2 (U - \tilde \Delta_k))}{\tilde \Delta_k^2 (U - \tilde \Delta_k)}.
\end{equation}
Since the feature is located roughly halfway to the TLS frequency, it is in principle possible to detect TLSs with frequencies outside the usable frequency range of the drive. However, the oscillation frequency of the $\ket{11}$ transition is significantly slower than the $\ket{01}$ transition, and decoherence can make the oscillation undetectable.

\subsection{Dissipation and dephasing}\label{sec:decoherent}
\begin{figure}
    \centering
    \includegraphics[width = \columnwidth]{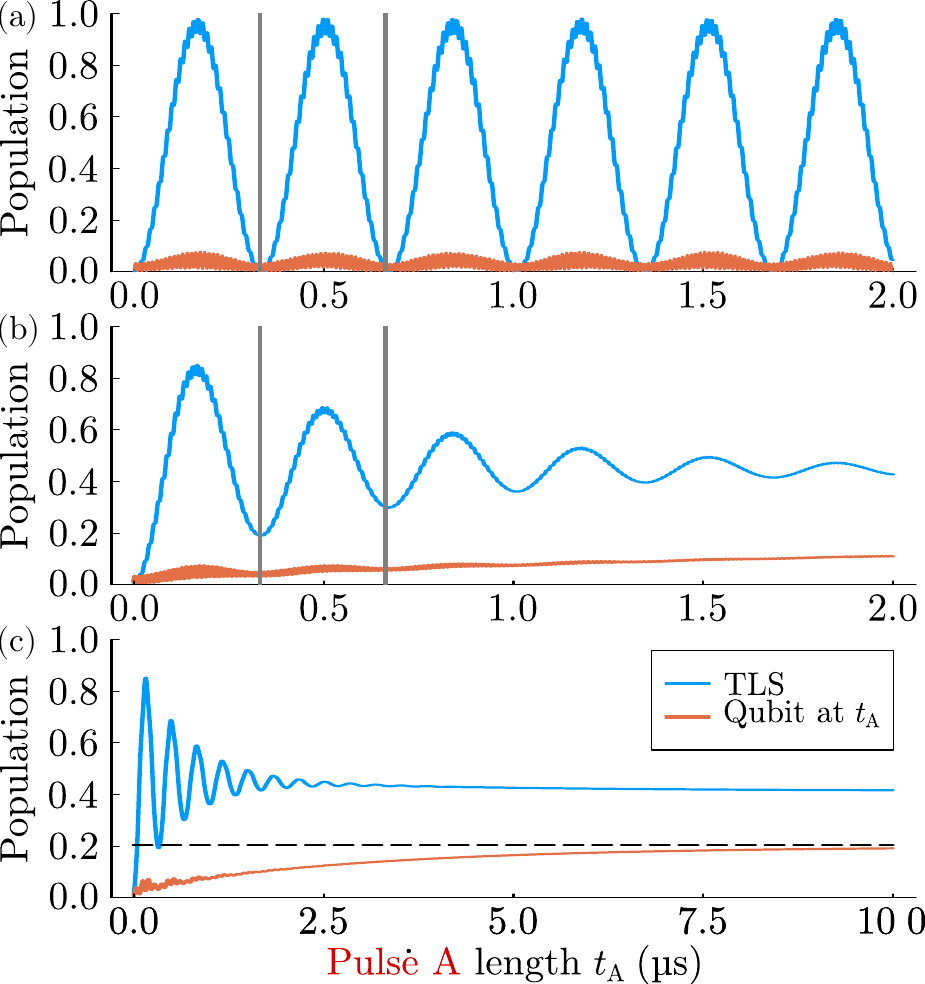}
    \caption{The populations $P_k(t_A) = \bra{\psi(t_A)} (\ket{01}\bra{01}) \ket{\psi(t_A)}$ of the TLS (blue) and transmon $P_q(t_A)$ (green) as a function of the length $t_A$ of pulse A, when the transmon is driven at the TLS frequency. The time-evolution is simulated without decoherence in (a) and with in (b) and (c). The vertical lines in (a) and (b) mark the times $t = 2 \pi \tilde \Delta_k / (A g_k)$ and $t = 4 \pi \tilde \Delta_k / (A g_k)$. The dashed horizontal line in (c) marks the steady-state transmon population $P_{q}(t_A \to \infty)$ predicted by Eq.~\eqref{eq:steady}}
    \label{fig:fig3}
\end{figure}

Let us next see what dissipation and dephasing affect the features described in the previous section. In Fig.~\ref{fig:fig3}, we compare the time-evolution of the TLS and transmon populations with and without decoherence. Clearly decoherence has no effect on the drive frequencies of the transitions. There is also no difference in the Rabi frequencies $\Omega_{01}$ and $\Omega_{11}$, and although the oscillation decays as one would expect, the frequency is still readily readable from the ($\omega, t$)-map.

It can also be of interest to know the dissipation and dephasing times $T_1$ and $T_2$ of the TLS. One can, of course, measure the dissipation time by exciting the TLS by driving it at the frequency gleaned from the $(\omega, t)$-map, and probing its population with the transmon frequency pulse B as a function of the delay between the pulses. Even though the $\pi$-pulse probe cannot be used to measure the TLS population exactly, one should be able to fit a reasonably accurate exponential decay curve. The dephasing time should be also obtainable via a Ramsey-type experiment. 

Alternatively, one can look at the steady state population of the transmon when it is driven with the frequency of a TLS. The time-evolution of the transmon-TLS system works on three distinct timescales. The fastest of these is the Rabi oscillation due to the drive. Another timescale is set by the decoherence time. Beyond this, decoherence has reduced the density matrix to a diagonal form, and its time-evolution can be described with a rate equation~\cite{busel_dissipation_2023}. The steady state of the rate equation uniquely determined by the Rabi frequency $\Omega_{01}$, and the dissipation and dephasing rates of the transmon-TLS pair. For the $\ket{00}$ to $\ket{01}$ transition, the transmon population in the steady state without pulse B is given by 
\begin{equation}\label{eq:steady}
    P_{q}(t_A \to \infty) = \frac{2 g^2 \Gamma + A^2 (\gamma_q + \kappa_q)}{6 g^2 \Gamma + 3 A^2 (\gamma_q + \kappa_q) + 8 \gamma_q \Delta^2},
\end{equation}
where $\Gamma = \gamma_q + \gamma_k + \kappa_q + \kappa_k$. See App.~\ref{app:rate} for more detail on how one can arrive at this result.  

\subsection{Overlapping features}\label{sec:ML}
When two or more TLSs have similar transition frequencies $\omega_k$ to each other or the second excited level of the transmon, the above analytical expressions no longer work, and estimating the TLS parameters becomes more difficult. Instead of trying to derive further analytic expressions for every possible case, we demonstrate that convolutional neural networks offer an accurate and efficient alternative. Ideas we develop here are similar to the ones exposed in ~\cite{niu_learning_2019}. 
\par
\begin{figure}[h]
    \centering
    \subfigure[]{
    \includegraphics[width=9cm]{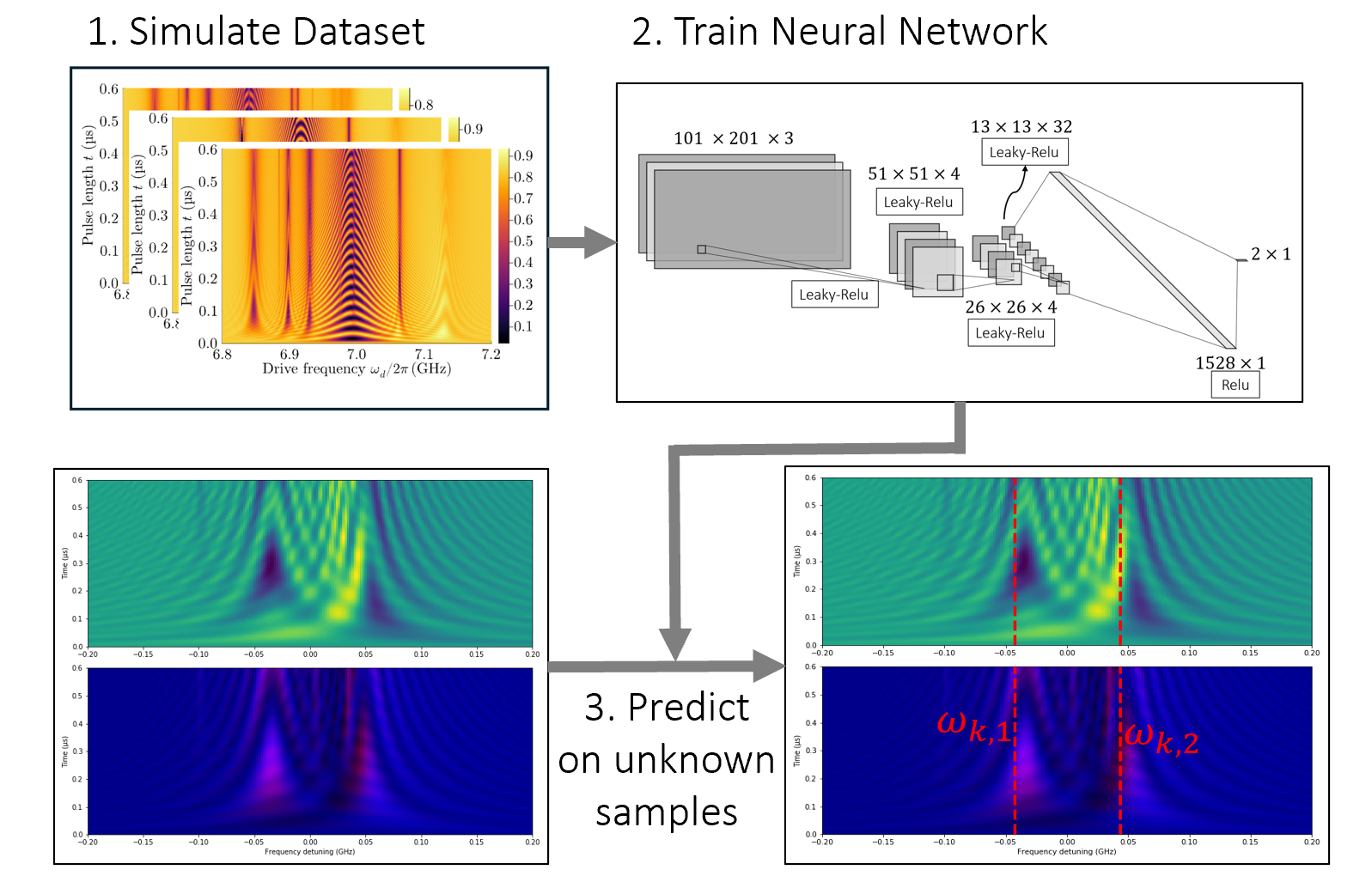}
    \label{fig:Arch}}
    \subfigure[]{
    \includegraphics[width=9cm]{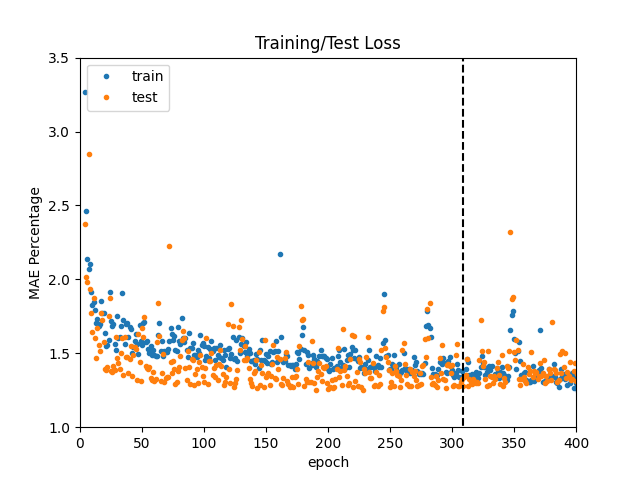}
    \label{fig:Loss}}
    \caption{(a) Schematic of workflow for training and implementation of convolutional neural network for determing TLS frequencies. (b) Training/test loss, dashed lines marks location of early stopping based on test loss.}
    \label{fig:ML}
\end{figure}

 
To benchmark this approach we consider the case of a two TLS system. Varying the parameters of the constituent TLSs, we generated training data set of 3800 $(\omega,t)$-maps labeled by the frequencies of the two TLSs. We then divide the data into an 80/10/10 split for train/test/validation data. We then construct a convolutional neural network for which the architecture is shown in Fig. \eqref{fig:Arch}. Implementing early stopping based on the mean absolute error percentage for the test data, the model realizes $98.6\%/98.7\%/98.6\%$ accuracy for the train/test/validation sets respectively. The training loss is shown in Fig. \eqref{fig:Loss}. 
\par
Noting that the analytical expressions are based on perturbation theory, it is very likely that the convolutional neural network predicts the parameters with greater accuracy even with the limited training set. Generating a larger set with up to three TLSs per map is fairly easy. However, for the network to be applicable to any experimental data, we would need to either vary the transmon parameters $\omega_q$ and $U$ and the drive amplitudes $A$, or tailor the generated data to a specific experimental device.

\section{Conclusions}\label{sec:conclusions}
In this work, we have introduced a two-tone spectroscopy to probe the TLS that are strongly coupled transmon. We propose a protocol for detecting TLSs in fixed-frequency qubits using a sequence of two microwave pulses. Pulse A is for exciting the TLSs and the pulse B at the transmon frequency is for detecting change in the TLS population. The parameters of the TLSs can be extracted from the information contained in the ($\omega, t$)-map of the transmon population as a function of the drive frequency and the length of pulse A. We have shown that it is simple to exctract the TLS frequencies $\omega_k$ from the locations of the chevron patterns corresponding to the cross-resonance-type oscillation of a given TLS on the frequency axis of the maps. The coupling strengths $g_k$ can also be estimated from the oscillation frequency of the same pattern. The same parameters are also possible to estimate from the feature corresponding to the bSWAP-type oscillation, although due to its much lower oscillation frequency, computing the parameters may be hindered by decoherence. With a high enough drive amplitude $A$, at least the cross-resonance oscillation should be observable even with relatively short dissipation and dephasing times. We also point out that the dissipation and pure dephasing times $T_{1, k}$ and $T_{\varphi, k}$ can be estimated from the steady state population of the qubit. We should note that the method is limited by the ratio of the coupling strength of a TLS to its frequency detuning $g_k/\Delta_k$ in a few ways. If it is too small, the shift of Eq.~\eqref{eq:shift} in the transmon frequency is too small to detect via pulse B. Since the Rabi frequency $\Omega_{01}$ of the $\ket{01}$ transition depends on the ratio, if the TLS dissipation time $T_{1_k}$ is too short, the TLS population may never grow large enough to be detectable. Additionally, for the Rabi frequency $\Omega_{01}$ to be measurable, the dephasing time of the TLS $T_{2, k}$ needs to be longer than the period of the oscillation.

It has been shown that TLSs frequencies can be tuned by applying strain~\cite{grabovskij_strain_2012, lisenfeld_decoherence_2016} or an external electric field~\cite{bilmes_resolving_2020}. With the ability to tune the TLS frequency, the TLS parameters frequencies and couplings could be found without the need to vary the length of pulse A.

We find that convolutional neural networks can be used to efficiently and accurately estimate the parameters, even when it is otherwise difficult due to overlapping chevron patterns in the ($\omega, t$)-maps. While the example shown in this work is quite simple, and only estimates TLS frequencies, we expect that the other parameters can be similarly estimated. Of course, for the neural networks to have any practical application, the training data needs to include varying qubit parameters as well as those of the TLSs. Construct such a network is a subject of future work. We should note that applying machine learning techniques to TLS spectroscopy is not an original idea~\cite{niu_learning_2019}. 

Devices with fixed-frequency qubits as their building blocks require microwave pulses for their two-qubit gates in addition the single-qubit gates. Since the pulses can result in the TLSs gaining population, it may very well be that fixed-frequency qubits are more susceptible to the degradation of gate fidelities over prolonged operation times. It is therefore useful to have a procedure for finding whether there are TLSs that may interfere with microwave gates due to leakage to the TLSs and the resulting shifts in the qubit frequencies. The information may help in the calibration of pulse sequences to minimize the effect of the TLSs on the gate fidelities. More generally, data on individual TLSs in fixed-frequency qubits may very well help in elucidating the physical origin of some of the TLSs. At the least by exposing differences in the TLSs of fixed-frequency qubits and their tunable counterparts. We leave the more detailed analysis of CNN approach to ($\omega, t$)-map analysis for the separate investigation. 

\section{Acknowledgements}
We are grateful to team at Google, Y. Chen,  A. Korotkov, P. Krogstrup, D. Kuzmanovski, J. Pekola, P. Roushan, V. Smilyansky, and S. Girvin for useful discussions. This work was supported by the European Research Council ERC HERO-810451 synergy grant, KAW foundation 2019-0068 and the Swedish Research Council (VR). Work at University of Connecticut was supported by the OVPR Quantum CT. 
\appendix

\section{Steady state transmon population}\label{app:rate}
The long timescale evolution of the population of a transmon-TLS pair can be approximated with the rate equations~\cite{busel_dissipation_2023, spiecker_two-level_2023}
\begin{align}
    \dot P_{0} = &\gamma_q P_{q} + \gamma_k P_{\tls} + c_1 (P_{q} - P_{0}) \\ \nonumber
    &+ c_2 (P_{\tls} - P_{0}) \\ \nonumber
    \dot P_{q} = &-\gamma_q P_{q} + c_1 (P_{0} - P_{q}) \\ \nonumber
    &+ c_3 (P_{\tls} - P_{q}) \\ \nonumber
    \dot P_{\tls} &= -\gamma_k P_{\tls} + c_2 (P_{0} - P_{\tls})\\ \nonumber
    &+ c_3 (P_{q} - P_{\tls}), \\ \nonumber
\end{align}
where 
\begin{equation}
    c_1 = \frac{(A/2)^2 (\gamma_q + \kappa_q)}{\Delta^2 + (\gamma_q + \kappa_q)^2},
\end{equation}
\begin{equation}
    c_2 = \frac{\Omega_{01}^2 (\gamma_k + \kappa_k)}{(2 g^2 / \Delta)^2 + (\gamma_k + \kappa_k)^2},
\end{equation}
and
\begin{equation}
    c_3 = \frac{g^2 \Gamma}{\Delta^2 + \Gamma^2}.
\end{equation}
Here we have assumed that $g, A \ll \Delta$, and that the dissipation is fast enough that the total population of the pair never exceeds one. The above equations give the steady state transmon population
\begin{align}
    &P_q(t \to \infty) = \\ \nonumber
    &\frac{c_2 c_3 + c_1 (c_2 + c_3 + \gamma_k)}{3 c_1 (c_2 + c_3) + 2 c_1 \gamma_k + \gamma_q (2 c_2 + \gamma_k) + c_2 (3 c_3 + \gamma_q + \gamma_k)},\\ \nonumber
\end{align}
which can be further approximated as
\begin{equation}
    P_q(t \to \infty) \approx \frac{2 g^2 \Gamma + A^2 (\gamma_q + \kappa_q)}{6 g^2 \Gamma + 3 A^2 (\gamma_q + \kappa_q) + 8 \gamma_q \Delta^2}.
\end{equation}

\section{Numerical details}
The Lindblad master equation of Eq.~\eqref{eq:lindblad} can be written in terms of the Liouvillian superoperator $\mathcal{L}(t)$ as
\begin{equation}
    \frac{d \dens}{d t} = \mathcal{L}(t) \dens.
\end{equation}
The solution to this can be written in terms of the matrix-vector product~\cite{am-shallem_three_2015}
\begin{equation}\label{eq:evo}
    \bar \rho(t) = e^{t L(t)} \bar \rho(0),
\end{equation}
where $\bar \rho$ is the vectorized density operator constructed by arranging the columns of the $d \times d$ matrix as a $1 \times d^2$ vector. The matrix representation of the Liouvillian superoperator $L(t)$ can be constructed by replacing the three-matrix products $\hat O_1 \dens \hat O_2$ by 
$((O_2^\dagger)^T \otimes O_1) \bar \rho$, where $O_1$ and $O_2$ are the $d \times d$ matrix representations of the operators $\hat O_1$ and $\hat O_2$. Since our Transmon-TLS systems have a relatively small Hilbert spaces, solving the time-evolution Eq.~\eqref{eq:evo} by exact diagonalization is fairly fast when the Liouvillian superoperator is time-independent. To get rid of the time-dependence of the drive Hamiltonian of Eq.~\eqref{eq:hamd}, we use the rotating-wave approximation for both pulses separately. The pulse B qubit population is computed as
\begin{equation}
    P(\omega_d, t_A = N \delta t) = \bar P_q^\dagger e^{t_\pi L_B} (e^{\delta t L_A(\omega_d)})^N \bar \rho_0,
\end{equation}
where $\bar \rho_0$ is the vectorized initial state, $\bar P_q$ is the vectorized qubit population operator, $L_A$ is the Liouvillian describing pulse A, and $L_B$ pulse B . The matrix-vector product $\bar P_q^\dagger e^{t_\pi L_B}$ needs to be computed only once for a set of transmon and TLS parameters, and $e^{\delta t L_A(\omega_d)} \bar \rho_0$ once for each drive frequency $\omega_d$. This reduces the computation of an $(\omega, t)$-map to a single matrix-vector product and an inner product for each $(\omega, t)$-pair. The numerical simulations were written in the Julia programming language~\cite{bezanson_julia_2017}. 

The example shown in Fig.~\ref{fig:fig2} was computed using the transmon parameters $\omega_q / 2\pi = \SI{7}{\giga\hertz}$, $U = \SI{180}{\mega\hertz}$, $T_{1, t} = \SI{10}{\micro\second}$ and $T_{\varphi, t} = \SI{1}{\micro\second}$. The drive amplitude for pulse B was $A / 2\pi = 1/(\SI{200}{ns}) = \SI{5}{\mega\hertz}$, meaning a $\pi$ rotation in the transmon state is achieved in $\SI{100}{ns}$. Pulse A amplitude was $A / 2 \pi = 1/\SI{60}{\mega\hertz}$. The TLS parameters are $\omega_k / 2\pi = \SI{7.08}{\giga\hertz}$, $g_k = \SI{30}{\mega\hertz}$, $T_{1, k} = \SI{800}{\nano\second}$ and $T_{\varphi, k} = \SI{1.6}{\micro\second}$. For Fig.~\ref{fig:fig3}, the coupling strength was $g_k = \SI{30}{\mega\hertz}$, otherwise the parameters are the same. For the training data used in Sec.~\ref{sec:ML}, the $\omega_k / 2\pi \in [\SI{6.8}{\giga\hertz}, \SI{7.2}{\giga\hertz}]$, $g_k \in [\SI{5}{\mega\hertz}, \SI{50}{\mega\hertz}]$,  $T_1\in [\SI{0.5}{\micro\second}, \SI{10}{\micro\second}]$, and $T_{\varphi, k} \in [\SI{0.5}{\micro\second}, \SI{30}{\micro\second}]$. Each one is randomly picked from a uniform distribution. We neglected the direct effect of the drive on the TLSs in all numerical simulations by setting $\lambda = 0$.

\bibliography{main}
\end{document}